%% file: longmore_antfest.tex
\documentclass[graybox]{svmult}

\usepackage{mathptmx}       
\usepackage{helvet}         
\usepackage{courier}        
\usepackage{type1cm}        
\usepackage{makeidx}         
\usepackage{graphicx}        
\usepackage{multicol}        
\usepackage[bottom]{footmisc}

\makeindex             

\begin{document}

\title*{Molecular gas in the inner 500pc of the Milky Way: violating
  star formation relations and on the verge of forming extreme stellar
  clusters}

\titlerunning{Gas in the CMZ: violating SF relations \& forming extreme clusters }

\author{Steven N. Longmore}

\institute{Karl-Schwarzschild-Str. 2, 85748 Garching bei M\"{u}nchen, Germany \email{slongmor@eso.org}}

%
%
\maketitle

\vspace{-2.5cm}

\abstract{ With the HOPS, MALT90 and HiGAL Galactic plane surveys we
  are mapping a significant fraction of the dense, star-forming,
  molecular gas in the Galaxy. I present results from two projects
  based on this combined dataset, namely, (i) looking for variations
  in the star formation (SF) rate across the Galaxy as a function of
  environment, and (ii) searching for molecular cloud progenitors of
  the most extreme (massive and dense) stellar clusters.  We find the
  SF rate per unit mass of \emph{dense} gas in the inner 500pc of the
  Galaxy is at least an order of magnitude lower than that in the
  disk, directly challenging the predictions of proposed universal
  column/volume density relations. In particular, the region
  $1^\circ<l<3.5^\circ$, $|b|<0.5^\circ$ contains
  $\sim$10$^7$\,M$_\odot$ of dense molecular gas --- enough to form
  1000 Orion-like clusters --- but the present-day star formation rate
  within this gas is only equivalent to that in Orion.  I present
  follow up studies of one molecular cloud we have studied as part of
  project (ii) which also lies in the inner 500\,pc of the Galaxy and
  is clearly extreme compared to the rest of the Galactic
  population. With a mass of 10$^5$ Msun, a radius of only $\sim$3pc
  and almost no signs of star formation it appears to be the
  progenitor of an Arches-like stellar cluster. Despite detailed
  observational follow-up searches, this object still appears to be
  unique in the Galaxy, making it extremely important for testing
  massive cluster formation models.}

\section{Introduction}
\label{sec:intro}

The conversion of gas into stars is one of the fundamental processes
in the Universe, and a cornerstone of astrophysics and cosmology.
Whether or not this process varies as a function of environment across
cosmological timescales underpins our understanding of the formation
of everything from planets to galaxy clusters. The key physics linking
all these fields is encapsulated in an end-to-end understanding of the
star and cluster formation process incorporating the effect of
different natal environmental conditions.

Developing an end-to-end model of star formation (SF) must be led by
observations that can both resolve sites of individual SF and
determine the global properties of their natal molecular clouds. For
the foreseeable future, SF regions in the Milky Way (MW) are the only
targets in the Universe for which this will be possible. The physical
environment across the MW varies widely, from the extreme conditions
close to the central supermassive black hole, to the benign conditions
in the outer Galaxy. By trying to understand what drives changes in
the stellar output of molecular clouds as a function of their physical
properties in our own Galaxy, we aim to provide a template for
understanding SF under similar conditions in external galaxies and
across cosmological timescales.

In many ways this is a Golden Age for SF studies in the MW. In the
next few years, the availability of multi-wavelength Galactic plane
survey data, new telescopes for follow-up studies, and advances in
numerical simulations will enable a revolution in Galactic SF
studies. Surveys are building GMC samples comprising a large fraction
of the dense molecular gas in the Galaxy. Combined with similar
surveys at IR and cm wavelengths, this will, for the first time,
provide statistically meaningful GMC samples separated by their
global/environmental properties and relative ages.

Within this context, I present some initial results from the
HOPS$^{1}$, MALT90$^2$ and HiGAL$^3$ Galactic plane surveys. We
conducted the simplest possible analysis of such an enormous dataset,
namely looking for i) large systematic variations between dense gas
tracers and SF indicators across the Galaxy, and ii) the most extreme
molecular clouds in the Galaxy.

\section{Testing star formation relations}
\label{sec:}

Recent surface- and volume-density star formation relations have been
proposed which potentially unify our understanding of how gas is
converted into stars, from the nearest star forming regions to
ultra-luminous infrared galaxies (ULIRGs)$^{4,5}$.  The inner 500\,pc
of our Galaxy contains the largest concentration of dense,
high-surface density molecular gas in the Milky Way$^{6,7}$, providing
an environment where the validity of these star-formation
prescriptions can be tested.

We have used recently-available data from HOPS and HiGAL at
wavelengths where the Galaxy is transparent, to find the dense,
star-forming molecular gas across the Milky Way$^8$. We use water and
methanol maser emission$^9$ to trace star formation activity within
the last $10^5$ years and 30 GHz radio continuum emission from the
Wilkinson Microwave Anisotropy Satellite (WMAP)$^{10}$ to estimate the
high-mass star formation rate averaged over the last $\sim4\times10^6$
years.

We find the dense gas distribution is dominated by the very bright and
spatially-extended emission within a few degrees of the Galactic
centre$^{11}$. This region accounts for $\sim$80\% of the NH$_3$(1,1)
integrated intensity but only contains 4\% of the survey
area. However, in stark contrast, the distribution of star formation
activity tracers is relatively uniform across the Galaxy.

To probe the dense gas vs SFR relationship towards the Galactic centre
region more quantitatively, we compared the HiGAL column density
maps$^{12}$ to the WMAP-derived SFR across the same region. The
results are shown in Figure 1. The total mass and SFR derived using
these methods agree well with previous values in the
literature$^{7,13,14}$. The main conclusion from this analysis is that
both the column-density threshold and volumetric SF relations$^{4,5}$
over-predict the SFR by an order of magnitude given the reservoir of
dense gas available to form stars. The region $1^\circ<l<3.5^\circ$,
$|b|<0.5^\circ$ is particular striking in this regard. It contains
$\sim$10$^7$\,M$_\odot$ of dense molecular gas --- enough to form 1000
Orion-like clusters --- but the present-day star formation rate within
this gas is only equivalent to that in Orion. This implication of this
result is that any universal column/volume density relations must be a
\emph{necessary but not sufficient} condition for SF to occur.

\section{Searching for molecular cloud progenitors of extreme stellar clusters}
\label{sec:}

Young massive clusters (YMCs) are thought to be the `missing link'
between open clusters and extreme extragalactic super star clusters
and globular clusters. We previously used the HOPS survey to search
for molecular clouds which may represent the initial conditions of
YMCs$^{15}$. With a mass of 10$^5$ M$_\odot$, a radius of only
$\sim$3pc and almost no signs of star formation, we put forward one
cloud, G0.253$+$0.016, as a likely progenitor of an Arches-like
YMC. Our Galactic plane survey data suggested this to be the most
massive and dense cloud in the Galaxy, making it extremely important
for testing massive cluster formation models$^{16}$.

In subsequent work we have attempted to quantify what the initial
conditions must be for molecular clouds to form bound YMCs, and
propose the progenitor clouds must have escape speeds greater than the
sound speed in photo-ionized gas$^{17}$. In these clumps, radiative
feedback in the form of gas ionization is bottled up, enabling star
formation to proceed to sufficiently high efficiency so that the
resulting star cluster remains bound even after gas removal. We
estimate the observable properties of clouds destined to form YMCs for
existing Galactic plane surveys.  Follow-up work searching through
BGPS data finds several clouds which pass the proposed
criteria$^{18}$. However, none of these are as massive as
G0.253$+$0.016 and in all of them prodigious star formation is already
underway.

Despite these, and other, detailed searches for similar objects,
G0.253$+$0.016 still appears to be the most massive and dense
molecular cloud in the Galaxy with almost no signs of star formation.

\begin{figure}
  \caption{Comparison of measured and predicted star formation rates
    as a function of Galacto-centric radius for the inner 500\,pc of
    the Milky Way. The colour scale shows the NH$_3$(1,1) integrated
    intensity image from HOPS$^{1}$ between $|l|\le3.5^\circ$ and
    $|b|\le0.5^\circ$ with the corresponding projected Galacto-centric
    radius underneath. The table below shows the observed and
    predicted gas mass and star formation rates for this region. The
    top two rows show the total gas mass and the mass of this gas
    above an extinction of A$_V=8$\,mag, respectively, derived from
    the HiGAL column density map$^{3}$. The third row shows the star
    formation rate derived from WMAP data$^{10}$. The fourth and fifth
    rows show the predicted star formation rates from Lada et al
    (2012)$^{4}$ and Krumholz, Dekel \& McKee (2012)$^{5}$
    respectively, given the gas mass in rows one and two. A volume
    density of 5$\times$10$^3$\,cm$^{-3}$ for gas in the CMZ was used
    to calculate the values in row 5$^{7}$. The different columns show
    the values for different longitude ranges, all with
    $|b|<0.5^\circ$. The second, third and fourth columns are for
    longitude ranges of $1^\circ<l<3.5^\circ$, $|l|<1^\circ$ and
    $-3.5^\circ<l<-1^\circ$, respectively. The final column shows the
    total across the full region of $|l|\le3.5^\circ$,
    $|b|\le0.5^\circ$.}
  \includegraphics[scale=.425]{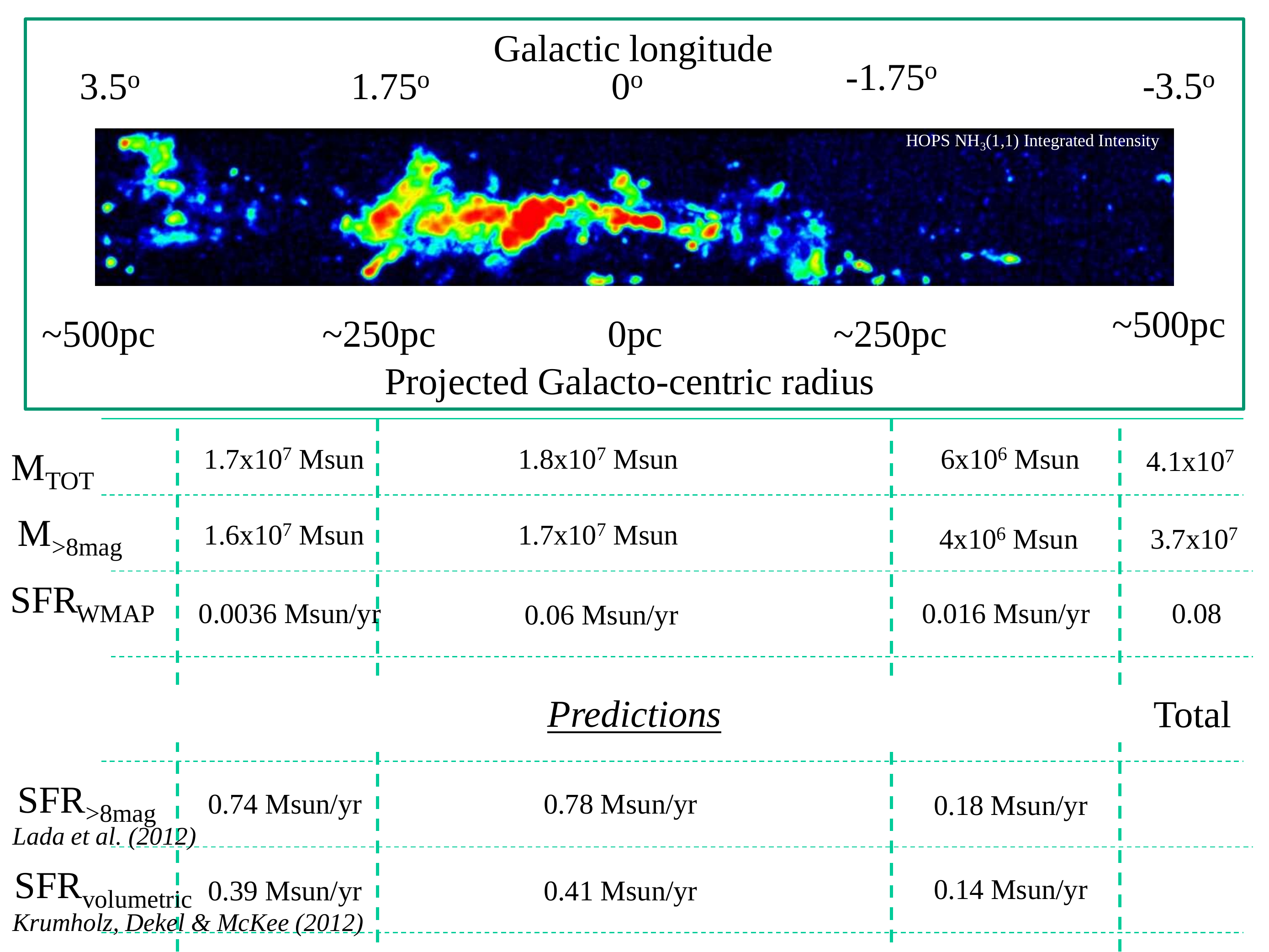}

\end{figure}

\input{longmore_antfest_reference}

\end{document}

%% file: longmore_antfest_reference.tex
%
%
%

%% file: longmore_antfest.bbl
\begin{thebibliography}{99.}%

\bibitem{phys-journal} Walsh, A. J. et al., 2011, MNRAS, 416, 1764-1821
\bibitem{phys-journal} Foster, J. et al., 2011, ApJS, 197, 25
\bibitem{phys-journal} Molinari, S. et al. 2010, A\&A, 518, 100-105 
\bibitem{phys-journal} Lada, C. J. et al., 2012, ApJ, 745, 190-196
\bibitem{phys-journal} Krumholz, M. et al.. 2012, ApJ, 745, 69-85
\bibitem{phys-journal} Morris, M. \& Serabyn, E., 1996,  ARAA, 34, 645-701
\bibitem{phys-journal} Ferriere, K. et al., 2007, A\&A, 467, 611-627
\bibitem{phys-journal} Longmore, S.N. et al, 2012, sub MNRAS, arXiv:1208.4256
\bibitem{phys-journal} Caswell, J. L. et al., 2010, MNRAS, 404, 1029-1060
\bibitem{phys-journal} Lee, E. et al., 2012, ApJ, 752, 146-160 
\bibitem{phys-journal} Purcell et al, 2012, MNRAS, 426, 1972
\bibitem{phys-journal} Molinari, S. et al., 2011, ApJ, 735, 33-40
\bibitem{phys-journal} Immer, K. et al., 2012, A\&A, 537, 121-140
\bibitem{phys-journal} Yusef-Zadeh, F. et al. 2009, ApJ, 702, 178-225
\bibitem{phys-journal} Longmore, S. N. et al., 2012, ApJ, 746, 117-127
\bibitem{phys-journal} Longmore, S. N. et al., 2011, ApJ, 726, 97
\bibitem{phys-journal} Bressert, E.  et al., 2012, ApJL, 758, 28
\bibitem{phys-journal} Ginsburg, A.  et al., 2012, ApJL, 758, 29

\end{thebibliography}
